\documentclass[twocolumn,conference]{IEEEtran}
\usepackage[T1]{fontenc}
\usepackage{verbatim}
\usepackage{float}
\usepackage{amsthm}
\usepackage{amsmath}
\usepackage{amssymb}
\usepackage{graphicx}
\usepackage[unicode=true,
 bookmarks=true,bookmarksnumbered=true,bookmarksopen=true,bookmarksopenlevel=1,
 breaklinks=false,pdfborder={0 0 0},backref=false,colorlinks=false]
 {hyperref}
\hypersetup{pdftitle={Your Title},
 pdfauthor={Your Name},
 pdfpagelayout=OneColumn, pdfnewwindow=true, pdfstartview=XYZ, plainpages=false}

\makeatletter

\floatstyle{ruled}
\newfloat{algorithm}{tbp}{loa}
\providecommand{\algorithmname}{Algorithm}
\floatname{algorithm}{\protect\algorithmname}

\theoremstyle{plain}
\newtheorem{thm}{\protect\theoremname}
\theoremstyle{plain}
\newtheorem{lem}[thm]{\protect\lemmaname}
\theoremstyle{remark}
\newtheorem{rem}[thm]{\protect\remarkname}

\usepackage[caption=false,font=footnotesize]{subfig}
\usepackage {algorithmic}
\usepackage{cite}

\makeatother

\providecommand{\lemmaname}{Lemma}
\providecommand{\remarkname}{Remark}
\providecommand{\theoremname}{Theorem}

\begin{document}

\title{Mean-Reverting Portfolio Design via Majorization-Minimization Method}

\author{\IEEEauthorblockN{Ziping~Zhao~and~Daniel~P.~Palomar}\IEEEauthorblockA{Department of Electronic and Computer Engineering\\
The Hong Kong University of Science and Technology, Hong Kong\\
Email: \{ziping.zhao, palomar\}@ust.hk}}

\maketitle
{\let\thefootnote\relax\footnotetext{This work was supported by the Hong Kong RGC 16206315 research grant.}}
\begin{abstract}
This paper considers the mean-reverting portfolio design problem arising
from statistical arbitrage in the financial markets. The problem is
formulated by optimizing a criterion characterizing the mean-reversion
strength of the portfolio and taking into consideration the variance
of the portfolio and an investment budget constraint at the same time.
An efficient algorithm based on the majorization-minimization (MM)
method is proposed to solve the problem. Numerical results show that
our proposed mean-reverting portfolio design method can significantly
outperform every underlying single spread and the benchmark method
in the literature.
\end{abstract}

\IEEEpeerreviewmaketitle{}

\section{Introduction}

Pairs trading \cite{Vidyamurthy2004,Ehrman2006,Bookstaber2011} is
a well known trading strategy which was pioneered by scientists Gerry
Bamberger and David Shaw, and the quantitative trading group led by
Nunzio Tartaglia at Morgan Stanley in the mid 1980s. As indicated
by the name, it is an investment strategy that focuses on a pair of
assets at the same time. Investors embracing this strategy do not
need to focus on the absolute future price of every single asset,
which by nature is hard to assess, but only the relative price of
this pair. To arbitrage from the market, investors need to buy the
under-priced asset, and short-sell the over-priced one and then profits
are locked in after trading positions are unwound when the relative
mispricing corrects itself in the future. In general, pairs trading
strategy with only two trading assets falls into the umbrella of statistical
arbitrage \cite{Pole2011,JacobsLevy2005,Nicholas2000} where the underlying
trading basket consists of three or more assets. 

In \cite{EngleGranger1987}, the authors first came up with the concept
of cointegration to describe the linear stationary relationship of
nonstationary time series which are said to be cointegrated. Later,
the cointegrated vector autoregressive model \cite{Johansen2000}
was proposed to discover cointegration relations which could be used
to get statistical arbitrage opportunities. Empirical and technical
analyses show that cointegration relations really exist in financial
markets, for example, the stock prices of the two American famous
consumer staple companies Coca-Cola and PepsiCo and those of the two
energy companies Ensco and Noble Corporation.%

Assets in a cointegration relation can be used to form a portfolio
and traded based on its stationary mean-reversion property. Such a
portfolio designed on a basket of assets is named as a mean-reverting
portfolio (MRP) which is also called a spread. In practice, spreads
are usually constructed using heuristic strategies and inherent correlations
may exist among them. However, they are commonly traded separately
with their possible connections neglected. So a natural question is
whether we can find an optimized portfolio of these spreads which
could outperform every single one. In this paper, this issue is precisely
addressed by designing a portfolio that combines multiple underlying
spreads.

In order to design a mean-reverting portfolio, there are two main
factors to consider: i) the designed MRP should exhibit a strong mean-reversion
indicating it can have frequent mean-crossing points and hence bring
in trading opportunities, and ii) the designed MRP should exhibit
enough variance meaning that it could lead to sufficient profits in
every trading and at the same time reduce the possibility that the
believed mean-reversion equilibrium breaks down. In \cite{dAspremont2011},
the author first gave insight on the MRP design problem without considering
the variance control. Later, authors in \cite{CuturidAspremont2013}
realized that directly solving the problem in \cite{dAspremont2011}
could result in a portfolio with very low variance, then the variance
control was taken into consideration and new criteria to characterize
the mean-reversion property were also proposed for the MRP design
problem. In \cite{CuturidAspremont2013}, semidefinite relaxation
methods were used to solve the nonconvex problem formulations, but
these methods are very computationally costly in general. Besides
that, the existing optimization problems in \cite{dAspremont2011,CuturidAspremont2013}
were all carried out by imposing an $\ell_{2}$-norm constraint on
the portfolio weight which is mathematically convenient but carries
no physical meaning in finance.

In this paper, we deal with the MRP design problem by optimizing a
mean-reversion criterion called portmanteau statistics, together with
the variance control and an investment budget constraint. An efficient
solving algorithm based on the MM method is proposed; this allows
the efficient resolution of the problem by a sequence of generalized
trust region subproblems. Numerical results show that our proposed
method significantly outperforms every single spread and the benchmark
method in \cite{CuturidAspremont2013} by comparing some performance
metrics through an designed trading experiment.

\section{Mean-Reverting Portfolio Design\label{sec:Mean-Reversion-Portfolio-Design}}

\subsection{Mean-Reverting Portfolio\label{sub:Mean-Reversion-Portfolio}}

For a financial asset like a common stock, its log-price at time $t$
is computed as $y_{t}=\log\left(p_{t}\right)$, where $\log\left(\cdot\right)$
is the natural logarithm, and $p_{t}$ denotes its price at time $t$.
For $M$ assets, their log-prices can be denoted as $\mathbf{y}_{t}=\left[y_{1,t},y_{2,t},\ldots,y_{M,t}\right]^{T}$.
An MRP is defined by the hedge ratio $\mathbf{w}_{s}$, and its corresponding
log-price spread $s_{t}$ is defined as 
\begin{equation}
s_{t}=\mathbf{w}_{s}^{T}\mathbf{y}_{t}=\sum_{m=1}^{M}w_{s,m}y_{m,t},\label{eq:spread}
\end{equation}
where $\mathbf{w}_{s}=\left[w_{s,1},w_{s,2},\ldots,w_{s,M}\right]^{T}$.
Vector $\mathbf{w}_{s}$ indicates the market value proportions invested
on the assets. For $m=1,2,\ldots,M$, $w_{s,m}>0$, $w_{s,m}<0$ and
$w_{s,m}=0$ mean a long position (i.e., the asset is bought), a short
position (i.e., the asset is short-sold), and no position, respectively. 

If we consider $N$ spreads, the log-prices for them can be denoted
as $\mathbf{s}_{t}=\left[s_{1,t},s_{2,t},\ldots,s_{N,t}\right]^{T}$.
Then an MRP can be designed on these spreads using weight $\mathbf{w}$
with its spread written as 
\begin{equation}
z_{t}=\mathbf{w}^{T}\mathbf{s}_{t}=\sum_{n=1}^{N}w_{n}s_{n,t},\label{eq:portfolio}
\end{equation}
where $\mathbf{w}=\left[w_{1},w_{2},\ldots,w_{N}\right]^{T}$ indicates
the market values on the spreads. Since we can get the centered counterpart
$\tilde{\mathbf{s}}_{t}=\mathbf{s}_{t}-\mathsf{E}\left[\mathbf{s}_{t}\right]$
for $\mathbf{s}_{t}$, we will use $\mathbf{s}_{t}$ to denote $\tilde{\mathbf{s}}_{t}$
in the following.

\subsection{Portmanteau Statistics $\mathrm{por}_{z}\left(p,\mathbf{w}\right)$\label{sub:Portmanteau-Statistics}}

The mean-reversion property for a time series $z_{t}$ is characterized
by the portmanteau statistics of order $p$ \cite{BoxTiao1977}. For
a centered univariate stationary process $z_{t}$, it is defined as
\begin{equation}
\mathrm{por}_{z}\left(p\right)\triangleq T\sum_{i=1}^{p}\rho_{i}^{2},
\end{equation}
where $\rho_{i}$ is the $i$-th order autocorrelation of $z_{t}$
defined by $\rho_{i}=\frac{\mathsf{E}\left[z_{t}z_{t-i}\right]}{\mathsf{E}\left[z_{t}^{2}\right]}$.
The portmanteau statistics is used to test whether a random process
is close to a white noise. From the definition, we have $\mathrm{por}_{z}\left(p\right)\geq0$
and the minimum of $\mathrm{por}_{z}\left(p\right)$ is attained by
a white noise process, i.e., the portmanteau statistics for a white
noise process is \textbf{$0$} for any $p$. Under this criterion,
in order to get a spread $z_{t}$ as close as possible to a white
noise process, we need to minimize $\text{por}_{z}\left(p\right)$
for a prespecified $p$. Given a spread $z_{t}=\mathbf{w}^{T}\mathbf{s}_{t}$,
since $\rho_{i}=\frac{\mathbf{w}^{T}\mathsf{E}\left[\mathbf{s}_{t}\mathbf{s}_{t+i}^{T}\right]\mathbf{w}}{\mathbf{w}^{T}\mathsf{E}\left[\mathbf{s}_{t}\mathbf{s}_{t}^{T}\right]\mathbf{w}}=\frac{\mathbf{w}^{T}\mathbf{M}_{i}\mathbf{w}}{\mathbf{w}^{T}\mathbf{M}_{0}\mathbf{w}}$
in which $\mathbf{M}_{i}$ is the $i$-th order autocovariance matrix
of $\mathbf{s}_{t}$, and specifically, when $i=0$, $\mathbf{M}_{0}$
stands for the (positive definite) covariance matrix, the $\mathrm{por}_{z}\left(p,\mathbf{w}\right)$
is written as
\begin{equation}
\mathrm{por}_{z}\left(p,\mathbf{w}\right)=T\sum_{i=1}^{p}\left(\frac{\mathbf{w}^{T}\mathbf{M}_{i}\mathbf{w}}{\mathbf{w}^{T}\mathbf{M}_{0}\mathbf{w}}\right)^{2}.\label{eq:portmanteau statistics for z}
\end{equation}

The matrices $\mathbf{M}_{i}$s in \eqref{eq:portmanteau statistics for z}
are assumed symmetric since they can always be symmetrized.

\subsection{MRP Design by Minimizing $\mathrm{por}_{z}\left(p,\mathbf{w}\right)$}

The MRP design problem can be formulated by minimizing the portmanteau
statistics $\mathrm{por}_{z}\left(p,\mathbf{w}\right)$ and at the
same time setting the variance of $z_{t}$ ($\mathsf{Var}\left[z_{t}\right]=\mathbf{w}^{T}\mathsf{E}\left[\mathbf{s}_{t}\mathbf{s}_{t}^{T}\right]\mathbf{w}=\mathbf{w}^{T}\mathbf{M}_{0}\mathbf{w}$)
to a certain level $\nu$ which could be written as follows:

\begin{equation}
\begin{array}{ll}
\underset{\mathbf{w}}{\mathsf{minimize}} & \sum_{i=1}^{p}\left(\mathbf{w}^{T}\mathbf{M}_{i}\mathbf{w}\right)^{2}\\
\mathsf{subject\:to} & \mathbf{w}^{T}\mathbf{M}_{0}\mathbf{w}=\nu\\
 & \mathbf{1}^{T}\mathbf{w}=1,
\end{array}\label{eq:problem formulation}
\end{equation}
where $\mathbf{w}^{T}\mathbf{M}_{0}\mathbf{w}=\nu$ is the spread
variance control with $\nu$ being a prespecified variance level and
$\mathbf{1}^{T}\mathbf{w}=1$ is the budget constraint which means
that all the money we invest on the long and short positions should
sum up to the investment budget. Here, the denominator in $\mathrm{por}_{z}\left(p,\mathbf{w}\right)$
is reduced because of the existence the variance constraint. In this
formulation, we allow short positions in the portfolio construction
which can enable the portfolio managers to achieve leverage by financing
portfolio not only with the budget but with the proceeds from the
assets held short positions. It is easy to see that the problem in
\eqref{eq:problem formulation} is nonconvex due to the nonconvexity
of the objective function and the constraint set.%

\section{Problem Solving Method via Majorization-Minimization\label{sec:Problem-Solving-Methods-MM}}

In this section, we first briefly discuss the majorization-minimization
or minorization-maximization (MM) method, and then the solving approach
for problem \eqref{eq:problem formulation} is given based on the
MM method.

\subsection{The MM Method}

The MM method \cite{HunterLange2004,RazaviyaynHongLuo2013,SunBabuPalomar2016}
refers to the majorization-minimization or minorization-maximization
method which is a generalization of the well-known expectation-maximization
(EM) algorithm. The idea behind MM is that instead of dealing with
the original optimization problem which could be difficult to tackle
directly, it solves a series of simple surrogate subproblems.

Suppose the optimization problem is as follows: 
\begin{equation}
\begin{array}{ll}
\underset{\mathbf{x}}{\mathsf{minimize}} & f\left(\mathbf{x}\right)\\
\mathsf{subject\:to} & \mathbf{x}\in{\cal X},
\end{array}
\end{equation}
where the constraint set ${\cal X}\subseteq\mathbb{R}^{N}$. In general,
there is no assumption on the convexity and differentiability of $f\left(\mathbf{x}\right)$.
The MM method aims to solve this problem by optimizing a sequence
of surrogate functions that majorize the objective function $f\left(\mathbf{x}\right)$
over the set ${\cal X}$. More specifically, starting from an initial
feasible point $\mathbf{x}^{\left(0\right)}$, the algorithm produces
a sequence $\left\{ \mathbf{x}^{\left(k\right)}\right\} $ according
to the following update rule:
\begin{equation}
\mathbf{x}^{\left(k+1\right)}\in\underset{\mathbf{x}\in{\cal X}}{\arg\min}\:u\left(\mathbf{x},\mathbf{x}^{\left(k\right)}\right),\label{eq:MM update}
\end{equation}
where $\mathbf{x}^{\left(k\right)}$ is the point generated by the
update rule at the $k$-th iteration and the surrogate function $u\left(\mathbf{x},\mathbf{x}^{\left(k\right)}\right)$
is the corresponding majorization function of $f\left(\mathbf{x}\right)$
at point $\mathbf{x}^{\left(k\right)}$. A surrogate function is called
a majorization function of $f\left(\mathbf{x}\right)$ at point $\mathbf{x}^{\left(k\right)}$
if it satisfies the following properties:
\begin{equation}
\begin{array}{cc}
u\left(\mathbf{x},\mathbf{x}^{\left(k\right)}\right)\geq f\left(\mathbf{x}\right), & \forall\mathbf{x}\in{\cal X},\\
u\left(\mathbf{x}^{\left(k\right)},\mathbf{x}^{\left(k\right)}\right)=f\left(\mathbf{x}^{\left(k\right)}\right).
\end{array}
\end{equation}
That is to say, the surrogate function $u\left(\mathbf{x},\mathbf{x}^{\left(k\right)}\right)$
should be an upper bound of the original function $f\left(\mathbf{x}\right)$
over ${\cal X}$ and coincide with $f\left(\mathbf{x}\right)$ at
point $\mathbf{x}^{\left(k\right)}$. Although the definition of $u\left(\mathbf{x},\mathbf{x}^{\left(k\right)}\right)$
gives us a lot of flexibility for choosing it, in practice, the surrogate
function $u\left(\mathbf{x},\mathbf{x}^{\left(k\right)}\right)$ should
be properly chosen so as to make the iterative update in \eqref{eq:MM update}
easy to compute. The MM method iteratively runs until some convergence
criterion is met. 

In the following, we will apply the MM method to the problem in \eqref{eq:problem formulation}.

\subsection{MRP Design Algorithm via MM}

To solve problem \eqref{eq:problem formulation} via majorization-minimization,
the key step is to find a majorization function of the objective function
such that the majorized subproblem is easy to solve. Observe that
the objective function is quartic in $\mathbf{w}$. The following
mathematical manipulations are necessary. We first compute the Cholesky
decomposition of $\mathbf{M}_{0}$ which is $\mathbf{M}_{0}=\mathbf{L}\mathbf{L}^{T}$,
where $\mathbf{L}$ is a lower triangular and nonsingular matrix.
Let us define $\bar{\mathbf{w}}=\mathbf{L}^{T}\mathbf{w}$, $\bar{\mathbf{M}}_{i}=\mathbf{L}^{-1}\mathbf{M}_{i}\left(\mathbf{L}^{T}\right)^{-1}$,
$\mathbf{c}=\mathbf{L}^{-1}\mathbf{1}$, and $\bar{\mathbf{W}}=\bar{\mathbf{w}}\bar{\mathbf{w}}^{T}$.
Then problem \eqref{eq:problem formulation} can be written as 

\begin{equation}
\begin{array}{ll}
\underset{\bar{\mathbf{w}},\bar{\mathbf{W}}}{\mathsf{minimize}} & \sum_{i=1}^{p}\left(\text{Tr}\left(\bar{\mathbf{M}}_{i}\bar{\mathbf{W}}\right)\right)^{2}\\
\mathsf{subject\:to} & \bar{\mathbf{W}}=\bar{\mathbf{w}}\bar{\mathbf{w}}^{T}\\
 & \bar{\mathbf{w}}^{T}\bar{\mathbf{w}}=\nu\\
 & \mathbf{c}^{T}\bar{\mathbf{w}}=1.
\end{array}\label{eq:problem formulation-1}
\end{equation}
Since $\text{Tr}\left(\bar{\mathbf{M}}_{i}\bar{\mathbf{W}}\right)=\text{vec}\left(\bar{\mathbf{M}}_{i}\right)^{T}\text{vec}\left(\bar{\mathbf{W}}\right)$,
where for matrix $\mathbf{M}$, $\text{vec}\left(\mathbf{M}\right)$
is a column vector consisting of all the columns of $\mathbf{M}$
stacked, then problem \eqref{eq:problem formulation} can be reformulated
as

\begin{equation}
\begin{array}{ll}
\underset{\bar{\mathbf{w}},\bar{\mathbf{W}}}{\mathsf{minimize}} & \text{vec}\left(\bar{\mathbf{W}}\right)^{T}\bar{\mathbf{M}}\text{vec}\left(\bar{\mathbf{W}}\right)\\
\mathsf{subject\:to} & \bar{\mathbf{W}}=\bar{\mathbf{w}}\bar{\mathbf{w}}^{T}\\
 & \bar{\mathbf{w}}^{T}\bar{\mathbf{w}}=\nu\\
 & \mathbf{c}^{T}\bar{\mathbf{w}}=1,
\end{array}\label{eq:problem formulation-2}
\end{equation}
where $\bar{\mathbf{M}}=\sum_{i=1}^{p}\text{vec}\left(\bar{\mathbf{M}}_{i}\right)\text{vec}\left(\bar{\mathbf{M}}_{i}\right)^{T}$.
Now, the objective function in \eqref{eq:problem formulation-2} becomes
a quadratic function of $\bar{\mathbf{W}}$, however, this problem
is still hard to solve. We then consider the application of the MM
trick on this problem \eqref{eq:problem formulation-2} by introducing
the following result.
\begin{lem}[{\cite[Lemma 1]{SongBabuPalomar2015}}]
\label{lem:Lemma majorization} Let $\mathbf{L}$ be a $K\times K$
symmetric matrix and $\mathbf{M}$ be another $K\times K$ symmetric
matrix such that $\mathbf{M}\succeq\mathbf{L}$. Then for any point
$\mathbf{x}_{0}\in\mathbb{R}^{K}$, the quadratic function \textbf{$\mathbf{x}^{T}\mathbf{L}\mathbf{x}$}
is majorized by $\mathbf{x}^{T}\mathbf{M}\mathbf{x}+2\mathbf{x}_{0}^{T}\left(\mathbf{L}-\mathbf{M}\right)\mathbf{x}+\mathbf{x}_{0}^{T}\left(\mathbf{M}-\mathbf{L}\right)\mathbf{x}_{0}$
at $\mathbf{x}_{0}$.
\end{lem}
According to Lemma \ref{lem:Lemma majorization}, given $\bar{\mathbf{W}}^{\left(k\right)}=\bar{\mathbf{w}}^{\left(k\right)}\bar{\mathbf{w}}^{\left(k\right)T}$
at iteration $k$, we know the objective function of problem \eqref{eq:problem formulation-2}
is majorized by the following majorization function at $\bar{\mathbf{W}}^{\left(k\right)}$:

\begin{equation}
\begin{array}{cl}
 & u\left(\bar{\mathbf{W}},\bar{\mathbf{W}}^{\left(k\right)}\right)\\
= & \psi\left(\bar{\mathbf{M}}\right)\text{vec}\left(\bar{\mathbf{W}}\right)^{T}\text{vec}\left(\bar{\mathbf{W}}\right)\\
 & +2\text{vec}\left(\bar{\mathbf{W}}^{\left(k\right)}\right)^{T}\left(\bar{\mathbf{M}}-\psi\left(\bar{\mathbf{M}}\right)\mathbf{I}\right)\text{vec}\left(\bar{\mathbf{W}}\right)\\
 & +\text{vec}\left(\bar{\mathbf{W}}^{\left(k\right)}\right)^{T}\left(\psi\left(\bar{\mathbf{M}}\right)\mathbf{I}-\bar{\mathbf{M}}\right)\text{vec}\left(\bar{\mathbf{W}}^{\left(k\right)}\right),
\end{array}\label{eq:majorization function for por}
\end{equation}
where $\psi\left(\bar{\mathbf{M}}\right)$ is a scalar depending on
$\bar{\mathbf{M}}$ and satisfying $\psi\left(\bar{\mathbf{M}}\right)\mathbf{I}\succeq\bar{\mathbf{M}}$.
Since the first term $\text{vec}\left(\bar{\mathbf{W}}\right)^{T}\text{vec}\left(\bar{\mathbf{W}}\right)=\left(\bar{\mathbf{w}}^{T}\bar{\mathbf{w}}\right)^{2}=\nu^{2}$
and the last term only depends on $\bar{\mathbf{W}}^{\left(k\right)}$,
they are just two constants. 
\begin{rem}
On the choosing of $\psi\left(\bar{\mathbf{M}}\right)$, according
to Lemma \ref{lem:Lemma majorization}, it is obvious that $\psi\left(\bar{\mathbf{M}}\right)$
can be easily chosen to be $\lambda_{max}\left(\bar{\mathbf{M}}\right)=\left\Vert \bar{\mathbf{M}}\right\Vert _{2}$.
In the implementation of the algorithm, although $\left\Vert \bar{\mathbf{M}}\right\Vert _{2}$
only need to be computed once for the whole algorithm, it is still
not computationally easy to get, so we can try to get an alternative
for $\left\Vert \bar{\mathbf{M}}\right\Vert _{2}$. Since for any
matrix $\mathbf{M}\in\mathbb{R}^{p\times q}$, we have $\left\Vert \mathbf{M}\right\Vert _{2}\leq\left\Vert \mathbf{M}\right\Vert _{F}=\sqrt{\sum_{i=1}^{p}\sum_{j=1}^{q}\left|m_{ij}\right|^{2}}$,
$\psi\left(\bar{\mathbf{M}}\right)$ can be chosen to be $\left\Vert \bar{\mathbf{M}}\right\Vert _{F}$
which is bigger than $\left\Vert \bar{\mathbf{M}}\right\Vert _{2}$
but much easier to compute.
\end{rem}
After ignoring the constants in \eqref{eq:majorization function for por},
the majorized problem of problem \eqref{eq:problem formulation-2}
is given by 

\begin{equation}
\begin{array}{ll}
\underset{\bar{\mathbf{w}},\bar{\mathbf{W}}}{\mathsf{minimize}} & \text{vec}\left(\bar{\mathbf{W}}^{\left(k\right)}\right)^{T}\left(\bar{\mathbf{M}}-\psi\left(\bar{\mathbf{M}}\right)\mathbf{I}\right)\text{vec}\left(\bar{\mathbf{W}}\right)\\
\mathsf{subject\:to} & \bar{\mathbf{W}}=\bar{\mathbf{w}}\bar{\mathbf{w}}^{T}\\
 & \bar{\mathbf{w}}^{T}\bar{\mathbf{w}}=\nu\\
 & \mathbf{c}^{T}\bar{\mathbf{w}}=1,
\end{array}\label{eq:problem formulation-MM}
\end{equation}
which can be further written as 

\begin{equation}
\begin{array}{ll}
\underset{\bar{\mathbf{w}},\bar{\mathbf{W}}}{\mathsf{minimize}} & \sum_{i=1}^{p}\text{Tr}\left(\bar{\mathbf{M}}_{i}\bar{\mathbf{W}}^{\left(k\right)}\right)\text{Tr}\left(\bar{\mathbf{M}}_{i}\bar{\mathbf{W}}\right)\\
 & -\psi\left(\bar{\mathbf{M}}\right)\text{Tr}\left(\bar{\mathbf{W}}^{\left(k\right)}\bar{\mathbf{W}}\right)\\
\mathsf{subject\:to} & \bar{\mathbf{W}}=\bar{\mathbf{w}}\bar{\mathbf{w}}^{T}\\
 & \bar{\mathbf{w}}^{T}\bar{\mathbf{w}}=\nu\\
 & \mathbf{c}^{T}\bar{\mathbf{w}}=1.
\end{array}\label{eq:problem formulation-MM-2}
\end{equation}

By changing $\bar{\mathbf{W}}$ back to $\bar{\mathbf{w}}$, problem
\eqref{eq:problem formulation-MM-2} becomes 
\begin{equation}
\begin{array}{ll}
\underset{\bar{\mathbf{w}}}{\mathsf{minimize}} & \bar{\mathbf{w}}^{T}\bar{\mathbf{H}}^{\left(k\right)}\bar{\mathbf{w}}\\
\mathsf{subject\:to} & \bar{\mathbf{w}}^{T}\bar{\mathbf{w}}=\nu\\
 & \mathbf{c}^{T}\bar{\mathbf{w}}=1,
\end{array}\label{eq:problem formulation-MM-3}
\end{equation}
where in the objective function, $\bar{\mathbf{H}}^{\left(k\right)}$
is defined in this way $\bar{\mathbf{H}}^{\left(k\right)}=\sum_{i=1}^{p}\left(\bar{\mathbf{w}}^{\left(k\right)T}\bar{\mathbf{M}}_{i}\bar{\mathbf{w}}^{\left(k\right)}\right)\bar{\mathbf{M}}_{i}-\psi\left(\bar{\mathbf{M}}\right)\bar{\mathbf{w}}^{\left(k\right)}\bar{\mathbf{w}}^{\left(k\right)T}$.
Finally, we can undo the change of variable $\bar{\mathbf{w}}=\mathbf{L}^{T}\mathbf{w}$,
obtaining 

\begin{equation}
\begin{array}{ll}
\underset{\mathbf{w}}{\mathsf{minimize}} & \mathbf{w}^{T}\mathbf{H}^{\left(k\right)}\mathbf{w}\\
\mathsf{subject\:to} & \mathbf{w}^{T}\mathbf{M}_{0}\mathbf{w}=\nu\\
 & \mathbf{1}^{T}\mathbf{w}=1,
\end{array}\label{eq:problem formulation-subproblem}
\end{equation}
where in the objective function, $\mathbf{H}^{\left(k\right)}=\sum_{i=1}^{p}\left(\mathbf{w}^{T\left(k\right)}\mathbf{M}_{i}\mathbf{w}^{\left(k\right)}\right)\mathbf{M}_{i}-\psi\left(\bar{\mathbf{M}}\right)\mathbf{M}_{0}\mathbf{w}^{\left(k\right)}\mathbf{w}^{\left(k\right)T}\mathbf{M}_{0}$. 

Observe that in the majorization problem \eqref{eq:problem formulation-subproblem},
the objective function becomes quadratic in the variable $\mathbf{w}$
rather than quartic as in the original problem \eqref{eq:problem formulation}.

\subsection{Solving the Majorization Problem in MM}

Thus, we just need to solve \eqref{eq:problem formulation-subproblem}
at every iteration. In order to solve \eqref{eq:problem formulation-subproblem},
one possible approach is via semidefinite programming (SDP). By using
matrix lifting $\mathbf{W}=\mathbf{w}\mathbf{w}^{T}$, the problem
\eqref{eq:problem formulation-subproblem} can be reformulated as

\begin{equation}
\begin{array}{ll}
\underset{\mathbf{W}}{\mathsf{minimize}} & \text{Tr}\left(\mathbf{H}^{\left(k\right)}\mathbf{W}\right)\\
\mathsf{subject\:to} & \text{Tr}\left(\mathbf{M}_{0}\mathbf{W}\right)=\nu\\
 & \text{Tr}\left(\mathbf{E}\mathbf{W}\right)=1\\
 & \mathbf{W}\succeq\mathbf{0},
\end{array}\label{eq:subproblem-eq01-MM-SDP}
\end{equation}
where $\mathbf{E}=\mathbf{1}\mathbf{1}^{T}$. The above SDP is empirically
observed to always give a rank one solution. However, since the computational
complexity of solving an SDP is high and it needs to be solved at
every iteration, it is not amenable for solving the problem \eqref{eq:problem formulation-subproblem}.

Here we introduce the solving approach by reformulating \eqref{eq:problem formulation-subproblem}
into a generalized trust region subproblem \cite{More1993}. Considering
$\mathbf{w}=\mathbf{w}_{0}+\mathbf{F}\mathbf{x}$ where $\mathbf{w}_{0}$
is any vector satisfying $\mathbf{1}^{T}\mathbf{w}_{0}=1$ and $\mathbf{F}$
is the kernel of $\mathbf{1}^{T}$ satisfying $\mathbf{1}^{T}\mathbf{F}=\mathbf{0}$
and a semi-unitary matrix satisfying $\mathbf{F}^{T}\mathbf{F}=\mathbf{I}$.
Let us define $\mathbf{N}^{\left(k\right)}=\mathbf{F}^{T}\mathbf{H}^{\left(k\right)}\mathbf{F}$,
$\mathbf{p}^{\left(k\right)}=\mathbf{F}^{T}\mathbf{H}^{\left(k\right)}\mathbf{w}_{0}$,
$b^{\left(k\right)}=\mathbf{w}_{0}^{T}\mathbf{H}^{\left(k\right)}\mathbf{w}_{0}$,
$\mathbf{N}_{0}=\mathbf{F}^{T}\mathbf{M}_{0}\mathbf{F}$ which is
positive definite, $\mathbf{p}_{0}=\mathbf{F}^{T}\mathbf{M}_{0}\mathbf{w}_{0}$,
and $b_{0}=\mathbf{w}_{0}^{T}\mathbf{M}_{0}\mathbf{w}_{0}$, then
the problem in \eqref{eq:problem formulation-subproblem} is equivalent
to

\begin{equation}
\begin{array}{ll}
\underset{\mathbf{x}}{\mathsf{minimize}} & \mathbf{x}^{T}\mathbf{N}^{\left(k\right)}\mathbf{x}+2\mathbf{p}^{\left(k\right)T}\mathbf{x}+b^{\left(k\right)}\\
\mathsf{subject\:to} & \mathbf{x}^{T}\mathbf{N}_{0}\mathbf{x}+2\mathbf{p}_{0}^{T}\mathbf{x}+b_{0}=\nu,
\end{array}\label{eq:GTRS}
\end{equation}
which is a nonconvex quadratically constrained quadratic programming.
Problems of this type \eqref{eq:GTRS} are called generalized trust
region subproblems (GTRS) \cite{More1993}. Such problems are usually
nonconvex but possess necessary and sufficient optimality conditions.
So efficient solving methods can be derived based on these conditions. 

According to Theorem 3.2 in \cite{More1993}, the optimality conditions
for the primal and dual variables $\left(\mathbf{x}^{\star},\xi^{\star}\right)$
of problem \eqref{eq:GTRS} are given as follows:

\begin{equation}
\begin{cases}
\left(\mathbf{N}+\xi^{\star}\mathbf{N}_{0}\right)\mathbf{x}^{\star}+\mathbf{p}+\xi^{\star}\mathbf{p}_{0}=0\\
\mathbf{x}^{\star T}\mathbf{N}_{0}\mathbf{x}^{\star}+2\mathbf{p}_{0}^{T}\mathbf{x}^{\star}+b_{0}-\nu=0\\
\mathbf{N}+\xi^{\star}\mathbf{N}_{0}\succeq\mathbf{0},
\end{cases}
\end{equation}
where the superscripts for the parameters are neglected hereafter.
We assume $\mathbf{N}+\xi\mathbf{N}_{0}\succ\mathbf{0}$, then we
can get the optimal solution is given by
\begin{equation}
\mathbf{x}\left(\xi\right)=-\left(\mathbf{N}+\xi\mathbf{N}_{0}\right)^{-1}\left(\mathbf{p}+\xi\mathbf{p}_{0}\right),
\end{equation}
and $\xi$ is the unique solution of the following equation with definition
on the interval ${\cal I}$ 
\begin{equation}
\begin{array}{cc}
\phi\left(\xi\right)=0, & \xi\in{\cal I}\end{array},
\end{equation}
where the function $\phi\left(\xi\right)$ is defined by 
\begin{equation}
\phi\left(\xi\right)=\mathbf{x}\left(\xi\right)^{T}\mathbf{N}_{0}\mathbf{x}\left(\xi\right)+2\mathbf{p}_{0}^{T}\mathbf{x}\left(\xi\right)+b_{0}-\nu,
\end{equation}
and the interval ${\cal I}$ consists of all $\xi$ for which $\mathbf{N}+\xi\mathbf{N}_{0}\succ\mathbf{0}$,
which implies that
\begin{equation}
{\cal I}=\left(-\lambda_{\min}\left(\mathbf{N},\mathbf{N}_{0}\right),\infty\right),
\end{equation}
where $\lambda_{min}\left(\mathbf{N},\mathbf{N}_{0}\right)$ is the
minimum generalized eigenvalue of matrix pair $\left(\mathbf{N},\mathbf{N}_{0}\right)$.
From Theorem 5.2 in \cite{More1993}, we know $\phi\left(\xi\right)$
is strictly decreasing on ${\cal I}$, then a simple line search algorithm
like bisection algorithm can be used to find the optimal $\xi$ over
${\cal I}$. In the beginning of the derivation, we assume $\mathbf{N}+\xi\mathbf{N}_{0}$
to be positive definite. This is a reasonable assumption since the
case when $\xi=-\lambda_{min}\left(\mathbf{N},\mathbf{N}_{0}\right)$
is very rare to happen theoretically and practically.

\subsection{Summary of the MM-Based Algorithm}

So, finally in order to solve the original problem \eqref{eq:problem formulation},
we just need to iteratively solve a sequence of GTRS problems in \eqref{eq:GTRS}.
In this paper, we call this MM-based algorithm iteratively reweighted
GTRS (IRGTRS). We summarize this algorithm in the following.

\begin{algorithm}[h]
\begin{algorithmic}[1] 
\REQUIRE $k=0$, $p$, $\nu$, and initial value $\mathbf{w}^{(0)}$.
\STATE Compute $\bar{\mathbf{M}}=\sum_{i=1}^{p}\text{vec}\left(\bar{\mathbf{M}}_{i}\right)\text{vec}\left(\bar{\mathbf{M}}_{i}\right)^{T}$ and $ \psi\left(\bar{\mathbf{M}}\right)$;
\REPEAT
\STATE Compute $\mathbf{H}^{\left(k\right)}=\sum_{i=1}^{p}\left(\mathbf{w}^{T\left(k\right)}\mathbf{M}_{i}\mathbf{w}^{\left(k\right)}\right)\mathbf{M}_{i}-\psi\left(\bar{\mathbf{M}}\right)\mathbf{M}_{0}\mathbf{w}^{\left(k\right)}\mathbf{w}^{\left(k\right)T}\mathbf{M}_{0}$;
\STATE Compute $\mathbf{w}^{\left(k+1\right)}$ by solving the GTRS problem \eqref{eq:GTRS};
\STATE $k=k+1$;
\UNTIL convergence 
\end{algorithmic} \protect\caption{IRGTRS - Iteratively Reweighted Generalized Trust Region Subproblem
algorithm for the MRP design problem in \eqref{eq:problem formulation}.}
\end{algorithm}

\section{Numerical Experiments\label{sec:Numerical-Experiments}}

In this section, we will conduct some numerical experiments to study
the performance of our proposed algorithm using synthetic data. 

We use a cointegrated system \cite{Luetkepohl2007} to generated $M=6$
time series for log-prices with cointegration rank $r=5$. Within
the $5$ cointegration relations, we estimate $N=r=5$ spreads. We
divide the sample path into $2$ overlapped sliding windows, and every
window is divided into two stages: in-sample training stage and out-of-sample
backtesting/trading stage. In the training stage, $T_{in}^{wnd}=12\times22=264$
samples corresponding to 12 months trading days are used to design
the MRP. In the trading stage, $T_{out}^{wnd}=6\times22=132$ samples
corresponding to 6 months trading days are used for simulated trading
experiments. In total, the length of the generated sample path is
$T=T_{in}^{wnd}+2\times T_{out}^{wnd}=528$. 

\begin{figure}[h]
\begin{centering}
\includegraphics[scale=0.63]{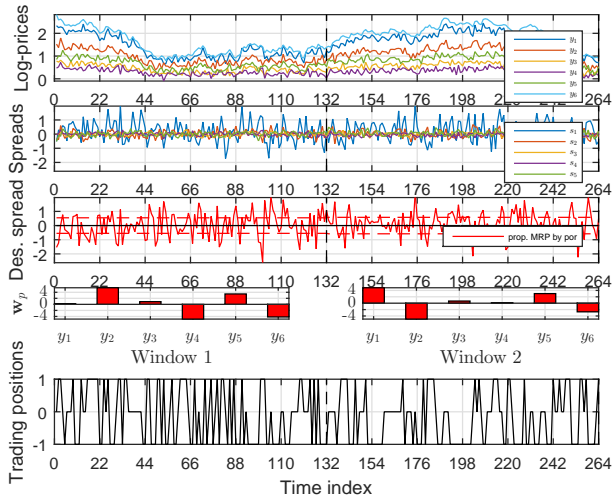}
\par\end{centering}

\protect\caption{\label{fig:trading-spread}Log-prices, spreads, a designed MRP using
portmanteau statistics $\mathrm{por}_{z}\left(\mathbf{w},3\right)$,
portfolio weights $\mathbf{w}_{p}$ (the weight on the underlying
assets) and trading positions for two trading windows. }
\end{figure}

We apply the designed spread $z_{t}$ to a mean-reversion trading
strategy. We first define a trading threshold $\delta$ for the spread.
Here we model $z_{t}$ as a Gaussian white noise, then according to
the result in \cite{Vidyamurthy2004}, the maximum trading profit
is attained when $\delta$ is set to be $0.75\times\mathsf{Std}\left(z_{t}\right)$,
where $\mathsf{Std}\left(\cdot\right)$ refers to the standard deviation.
Knowing the long-run equilibrium of $z_{t}$ to be $\mu_{z}$, a simple
trading framework is designed as follows: open a long position (denoted
as $1$), i.e., buy the portfolio $z_{t}$ at time $t_{1}$ when $z_{t_{1}}\leq\mu_{z}-\delta$
and unwind the position (denoted as $0$) if $\mu_{z}\leq z_{t_{1}+\tau_{1}}<\mu_{z}+\delta$
or change to a short position (denoted as $-1$) if $z_{t_{1}+\tau_{1}}\geq\mu_{z}+\delta$
at time $t_{1}+\tau_{1}$ which is the first time that $z_{t}$ satisfies
either of the above conditions; open a short position, i.e., short-sell
the portfolio $z_{t}$ at time $t_{2}$ if $z_{t_{2}}\geq\mu_{z}+\delta$
and unwind the position if $\mu_{z}-\delta<z_{t_{2}+\tau_{2}}\leq\mu_{z}$
or change to a long position if $z_{t_{2}+\tau_{2}}\leq\mu_{z}-\delta$
at time $t_{2}+\tau_{2}$ which is the first time that $z_{t}$ satisfies
either of the above conditions. Based on this trading scheme, one
example is shown in Figure \ref{fig:trading-spread}.

\begin{figure}[h]
\begin{centering}
\includegraphics[scale=0.63]{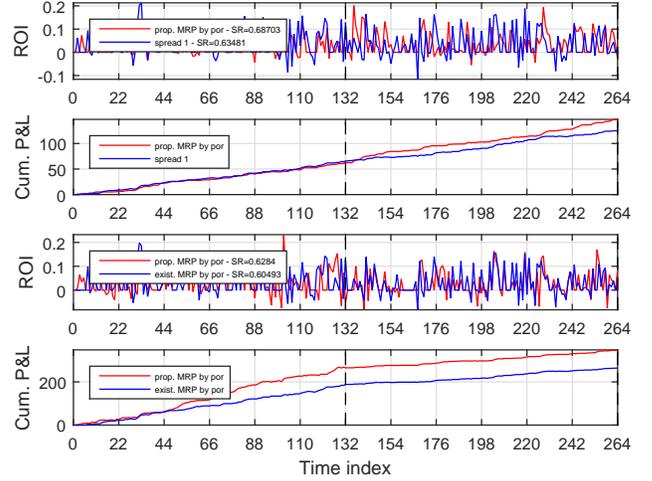}
\par\end{centering}

\protect\caption{\label{fig:performance-comparisons}Comparisons of ROIs, Sharpe ratios
of ROIs, and cumulative P\&Ls between the MRP designed using our proposed
method (prop. MRP by $\text{por}$) with one underlying spread (spread
1) and one benchmark method in \cite{CuturidAspremont2013} (exist.
MRP by $\text{por}$).}
\end{figure}

We also compare the cumulative profit and loss (P\&L), the return
on investment (ROI), and the Sharpe ratio of the ROI of our designed
portfolio with every underlying spread and the portfolio designed
using method in \cite{CuturidAspremont2013}. Cumulative P\&L measures
the return performance from the trading. ROI at time $t$ is defined
as the P\&L at time $t$ normalized by the gross investment deployed
(that is the gross exposure to the market including the long position
investment and the short position investment). In the computation
of Sharpe ratios, we set the risk-free return to $0$. Simulation
results show that our designed portfolio can outperform all the underlying
spreads and the design method in \cite{CuturidAspremont2013} with
a higher final cumulative P\&L and a higher Sharpe ratio. The comparison
results for two cases are given in Figure \ref{fig:performance-comparisons}. 

A more detailed discussion on this mean-reverting portfolio design
problem will be available in a journal version in preparation. 

\section{Conclusions}

The mean-reverting portfolio design problem arising from statistical
arbitrage has been considered in this paper. We have formulated the
problem by optimizing a mean-reversion criterion characterizing the
mean-reversion strength of the portfolio spread, and at the same time
taking into consideration the variance of the spread and an investment
budget constraint. Then an efficient algorithm based on MM method
has been proposed to solve the problem. Numerical results show that
our proposed method is able to generate consistent positive profits
and significantly outperform the underlying spreads and the design
method in the literature.

\appendices{}

\bibliographystyle{IEEEtran}
\bibliography{D:/Dropbox/Research/Reference/MRP}

\end{document}